# Physically Interpretable Descriptors Drive the Materials Design of Metal Hydrides for Hydrogen Storage

*Seong-Hoon Jang[*1], Di Zhang[1], Hung Ba Tran[1], Xue Jia[1], Kiyoe Konno[1, 2], Ryuhei Sato[3], Shin-ichi Orimo[*1, 4], and Hao Li[*1]*

[1] Advanced Institute for Materials Research (WPI-AIMR), Tohoku University, Sendai 980-8577, Japan

[2] Institute of Fluid Science, Tohoku University, Sendai, 980-8577, Japan

[3] Department of Materials Engineering, The University of Tokyo, Tokyo 113-8656, Japan

[4] Institute for Materials Research (IMR), Tohoku University, Sendai, 980-8577

*Corresponding authors:

jang.seonghoon.b4@tohoku.ac.jp (S.-H. Jang),

shin-ichi.orimo.a6@tohoku.ac.jp (S. Orimo),

li.hao.b8@tohoku.ac.jp (H. Li)

ABSTRACT

Designing metal hydrides for hydrogen storage remains a longstanding challenge due to the vast compositional space and complex structure-property relationships. Herein, for the first time, we present physically interpretable models for predicting two key performance metrics, gravimetric hydrogen density $w$ and equilibrium pressure $P_{\mathrm{eq,RT}}$ at room temperature, based on a minimal set of chemically meaningful descriptors. Using a rigorously curated dataset of 5,089 metal hydride compositions from our recently developed Digital Hydrogen Platform (*DigHyd*) based on large-scale data mining from available experimental literature of solid-state hydrogen storage materials, we systematically constructed over 1.6 million candidate models using combinations of scalar transformations and nonlinear link functions. The final closed-form models, derived from 2-3 descriptors each (*e.g.*, atomic mass, electronegativity, molar density, and ionic filling factor), achieve predictive accuracies on par with state-of-the-art machine learning methods, while maintaining full physical transparency. Strikingly, descriptor-based design maps generated from these models reveal a fundamental trade-off between $w$ and $P_{\mathrm{eq,RT}}$: saline-type hydrides, composed of light electropositive elements, offer high $w$ but low $P_{\mathrm{eq,RT}}$, whereas interstitial-type hydrides based on heavier electronegative transition metals show the opposite trend. Notably, beryllium (Be)-based systems, such as Be–Na alloys, emerge as rare candidates that simultaneously satisfy both performance metrics, attributed to the unique combination of light mass and high molar density for Be. Our models indicate that, while there remains room for improvement between the current state of solid-state hydrogen storage materials and the US-DOE targets, Be-based systems may offer renewed prospects for approaching these benchmarks. These results provide chemically intuitive guidelines for materials design and establish a scalable framework for the rational discovery of materials in complex chemical spaces. The methodology is broadly applicable and could serve as a template for data-driven exploration across other energy-relevant materials domains.

KEYWORDS. Metal hydrides, hydrogen storage materials, AI for Materials, interpretable model

## Introduction

Hydrogen is a leading candidate for enabling carbon-neutral energy technologies due to its high specific energy and clean combustion profile.[1, 2] However, its practical deployment in fuel cells and energy systems is constrained by the lack of compact, safe, and reversible storage solutions.[3] Among various strategies, solid-state hydrogen storage using metal hydrides has received significant attention owing to their high volumetric density, cyclability, and integrability into engineered systems.[4-6]

Metal hydrides, such as $MgH_2$, $Mg_2NiH_4$, $FeTiH_2$, $PdH_{0.6}$, and $LaNi_5H_6$, have long served as prototypical systems.[7-15] These materials span a wide thermodynamic range: saline hydrides, based on light metal atoms (*e.g.*, $MgH_2$) provide high gravimetric capacities but suffer from high decomposition temperatures,[13] while interstitial hydrides, based on transition or heavy metal atoms (*e.g.*, $LaNi_5H_6$) offer excellent kinetics and hydrogen equilibrium pressures but limited capacity.[11] Significant efforts have focused on modifying these systems through compositional tuning, nanostructuring, and catalysis to improve hydrogenation performance for the target metrics.[4-6]

Yet despite decades of study, the compositional landscape of hydride-forming alloys remains largely underexplored. Thousands of binary and multinary combinations are theoretically possible, yet only a small subset has been synthesized and evaluated. This data sparsity is compounded by the lack of predictive, physically grounded frameworks that can guide rational materials discovery. While recent machine learning (ML) efforts have shown potential in accelerating property prediction,[16, 17] they often rely on relatively small scale of data, poorly curated datasets, and opaque modeling strategies that limit interpretability and chemical insight.

To address these challenges, herein, we present a data-driven but physically interpretable approach to the design of metal hydrides. Using a rigorously curated dataset, our recently developed Digital Hydrogen Platform (*DigHyd*: www.dighyd.org) *via* large-scale data mining from available experimental literature on solid-state hydrogen storage materials, we construct regression models that accurately predict two key metrics: gravimetric hydrogen density ($w$) and equilibrium pressure ($P_{\mathrm{eq,RT}}$) at room temperature ($298.15$ K). Our models utilize a minimal set of intuitive descriptors (*i.e.*, atomic mass ($M$), electronegativity ($\chi$), molar density ($\rho_{\mathrm{mol}}$), and ionic filling factor ($\eta_f$)) and achieve predictive accuracy comparable to modern black-box ML algorithms. Importantly, these models enable the generation of compositional design maps, which reveal fundamental structure–property relationships and identify promising candidate systems—particularly beryllium (Be)-based alloys—for high-performance hydrogen storage. In doing so, they also highlight the challenges that remain in bridging the difference between the current performance of solid-state hydrogen storage materials and the ambitious targets set by the US-DOE, while indicating that Be-based systems may offer a viable pathway toward bridging this divide. This work provides a transparent and scalable framework for accelerating the rational discovery of solid-state hydrogen storage materials.

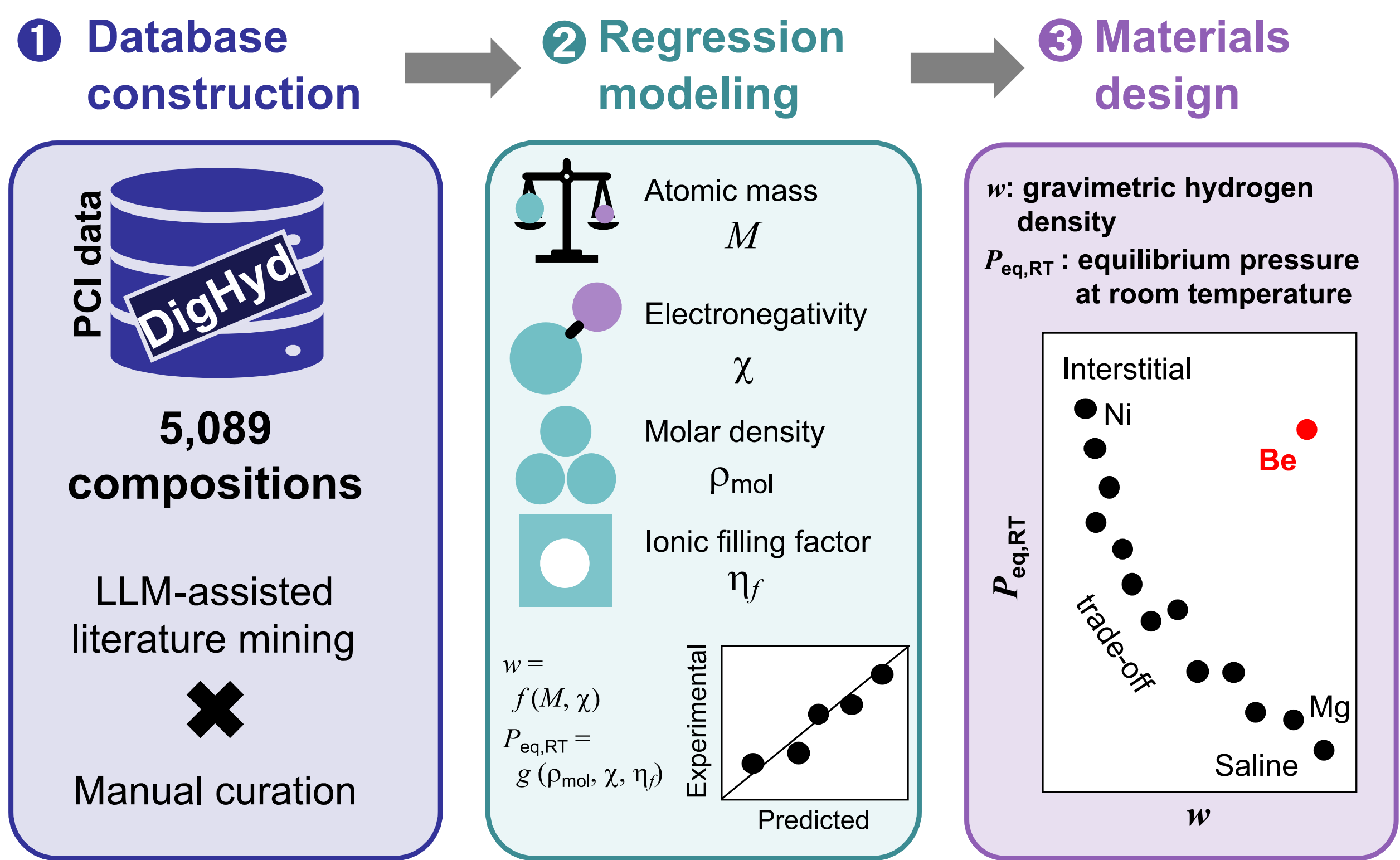

**Figure 1. Workflow for descriptor-based modeling and design of metal hydrides for hydrogen storage.** This workflow encompasses three stages: database construction, regression modeling, and materials design.

## Results and Discussion

A schematic overview of our descriptor-based approach to hydrogen storage materials discovery is presented in **Figure 1**. The workflow begins with the construction of a curated database, followed by regression modeling using physically meaningful descriptors, and culminates in compositional mapping for rational materials design. The database component is based on the *DigHyd* platform, which consolidates 5,089 pressure-composition isotherm (PCI) data points specifically for metal hydrides. These entries were initially identified from 910 literature sources using large language model assisted mining and then subjected to rigorous manual review. During this curation process, entries lacking clearly defined chemical formulae were excluded to ensure compositional consistency. For $P_{\mathrm{eq,RT}}$, we included values either directly reported in the literature or indirectly estimated using the van't Hoff equation when thermodynamic parameters, namely enthalpy ($\Delta H$) and entropy ($\Delta S$), were available from multi-temperature PCI data. After removing duplicate compositions, the final dataset comprised 1,967 and 1,078 unique entries for $w$ and $P_{\mathrm{eq,RT}}$, respectively, spanning a broad range of metal hydride chemistries.

While both *DigHyd* database and the previously developed ML_Hydpark[16] database represent valuable data resources for hydrogen storage research, *DigHyd*, which serves as a superset of ML_Hydpark, offers several distinct advantages for data-driven modeling. Most importantly, *DigHyd* includes an order of magnitude more PCI entries for metal hydrides (*DigHyd*: 5,089 *vs*. ML_Hydpark: 430)[16] and additionally encompasses covalent organic frameworks (COFs) and metal-organic frameworks (MOFs), which were not used in the present study but are available within the platform. Furthermore, *DigHyd* is structured in JavaScript Object Notation (JSON) format and includes full reference traceability *via* Digital Object Identifiers (DOIs), significantly improving data transparency and reproducibility. By contrast, ML_Hydpark is distributed in flat CSV format and lacks consistent bibliographic metadata, which limits its utility for model interpretation and downstream analysis. These attributes made *DigHyd* a more scalable and reliable foundation for building physically interpretable models in this work.

To develop low-cost, transparent predictive models, we found a set of effective descriptors that capture essential physical and chemical characteristics of metal hydride systems among the candidate features: $M$, $\chi$, $\rho_{\mathrm{mol}}$, and $\eta_f$. These variables serve as proxies for key mechanisms influencing hydrogen storage capacity and thermodynamic behavior: lattice weight, bond polarity, structural density, and steric occupancy. Closed-form regression models constructed from these descriptors were able to predict $w$ and $P_{\mathrm{eq,RT}}$ with high accuracy, while maintaining physical transparency, an essential attribute for guiding materials design beyond black-box ML prediction.

Finally, the model outputs were used to generate materials design maps that visualize hydrogen storage performance across a wide compositional space. Strikingly, these maps highlight an intrinsic trade-off between $w$ and $P_{\mathrm{eq,RT}}$: saline-type hydrides, such as Mg-based systems, tend to exhibit high $w$ but low $P_{\mathrm{eq,RT}}$, whereas interstitial-type hydrides, such as Ni-based systems, display the opposite trend. Notably, Be-based systems emerge as rare candidates that potentially circumvent this trade-off, achieving high $w$ and high $P_{\mathrm{eq,RT}}$. While the broader implications of this trend are explored in detail later, the present figure serves to frame the conceptual structure of this study: from curated data to interpretable modeling, to chemically meaningful design guidance (**Figure 1**).

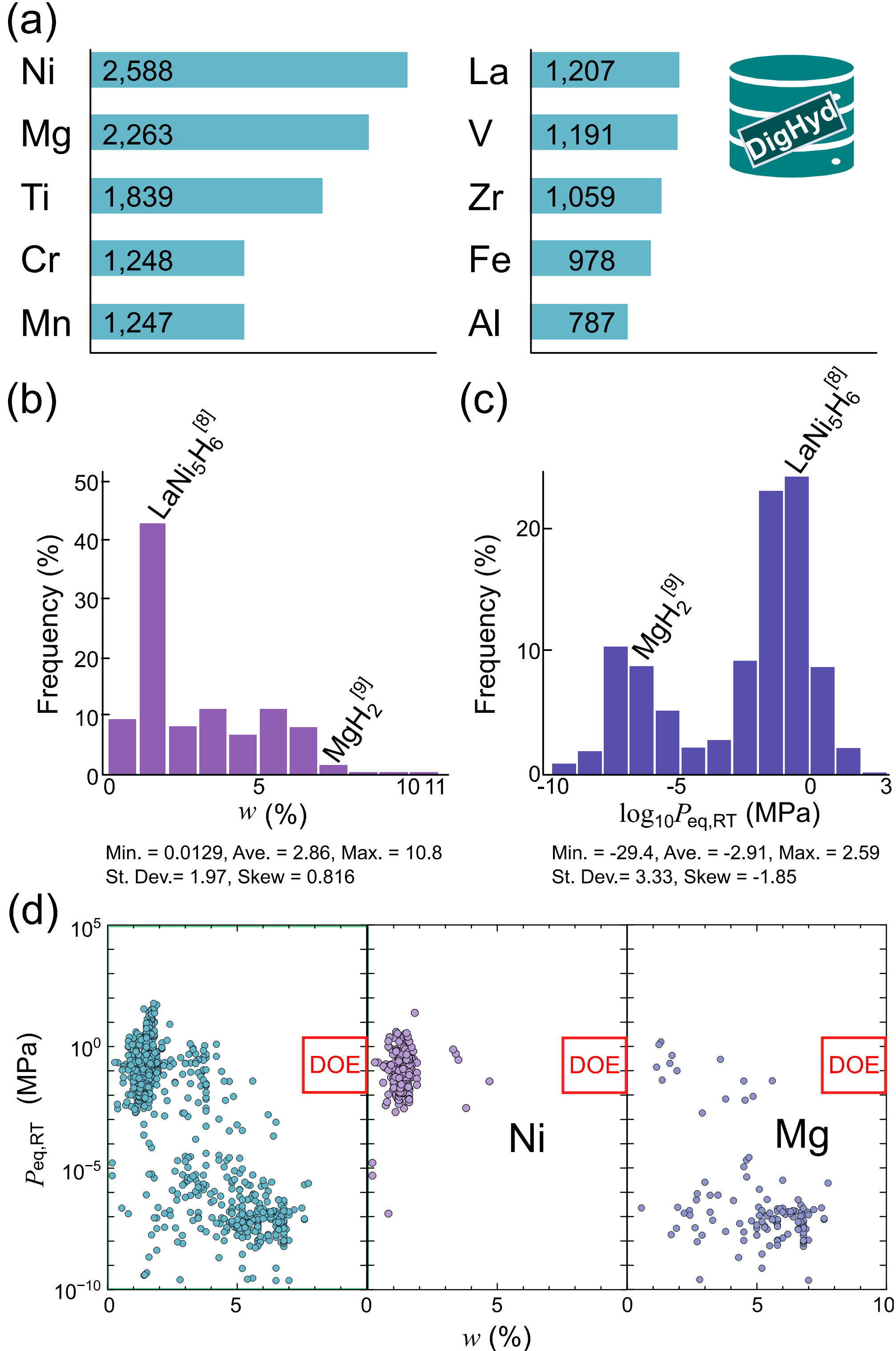

**Figure 2. Data profile of metal hydrides in the *DigHyd* database.** (a) Frequency of metal elements appearing in the dataset, with Ni and Mg being most prevalent. Histograms of (b) gravimetric hydrogen density ($w$) and (c) logarithm-scaled equilibrium pressure ($P_{eq,RT}$) at room

temperature, highlighting $MgH_2$ (saline-type) and $LaNi_5H_6$ (interstitial-type) as representative cases.[8, 9] For both (b) and (c), the minimum (Min.), average (Ave.), maximum (Max.), standard deviation (St. Dev.), and skewness (Skew) are provided, illustrating the broad spread and asymmetry in the dataset. (d) Scatter plots of $w$ versus $P_{\mathrm{eq,RT}}$: all compositions (left), Ni-containing but Mg-free systems (middle), and Mg-containing but Ni-free systems (right). The ultimate target region of U.S. Department of Energy (US-DOE) target region is indicated by red boxes.

To gain insights into the composition and diversity of the curated dataset, we analyzed the distribution of elements and key performance metrics within the *DigHyd* database. As shown in **Figure 2a**, certain metal elements appear more frequently than others, reflecting historical research focus and experimental accessibility. Ni (2,588 entries) and Mg (2,263 entries) dominate the dataset, consistent with the prevalence of interstitial- and saline-type hydrides, respectively. Ti, Cr, Mn, La, V, Zr, Fe, and Al also appear prominently, representing a range of compositional classes, which ensures that the dataset captures both conventional and less-explored regions of the hydride design space.

The distributions of the two target properties, $w$ and $P_{\mathrm{eq,RT}}$, are shown in **Figures 2b** and **2c**, respectively. The histogram of $w$ exhibits a right-skewed distribution, with most compositions clustering below 5%, and a long tail extending toward higher-capacity systems. While $MgH_2$, a classic saline-type hydride, lies on the higher-capacity end, $LaNi_5H_6$, a representative interstitial-type compound, sits near the modal value. Meanwhile, the distribution of $P_{\mathrm{eq,RT}}$ spans a wide range, highlighting the vast thermodynamic diversity of metal hydrides. Notably, $MgH_2$ and $LaNi_5H_6$ again serve as instructive references, occupying opposite extremes of the pressure-capacity landscape. The statistical summaries are also annotated beneath each plot, including the minimum, average, maximum, standard deviation, and skewness.

In addition to the elemental distributions and property histograms, **Figure 2d** provides a direct visualization of the relationship between $w$ and $P_{\mathrm{eq,RT}}$. For the full dataset (**Figure 2d, left**), a weak trade-off relationship between $w$ and $P_{\mathrm{eq,RT}}$ can be discerned. When focusing on Ni-containing but Mg-free systems, which are representative of interstitial-type hydrides (**Figure 2d, middle**), the trend shifts toward high $P_{\mathrm{eq,RT}}$ but low $w$. In contrast, Mg-containing but Ni-free systems, corresponding to saline-type hydrides (**Figure 2d, right**), display the opposite behavior, namely low $P_{\mathrm{eq,RT}}$ and high $w$. Notably, in all cases, the data points have not yet reached the US-DOE target region, outlined by red boxes, underscoring the substantial performance gap that persists in the current hydride landscape.

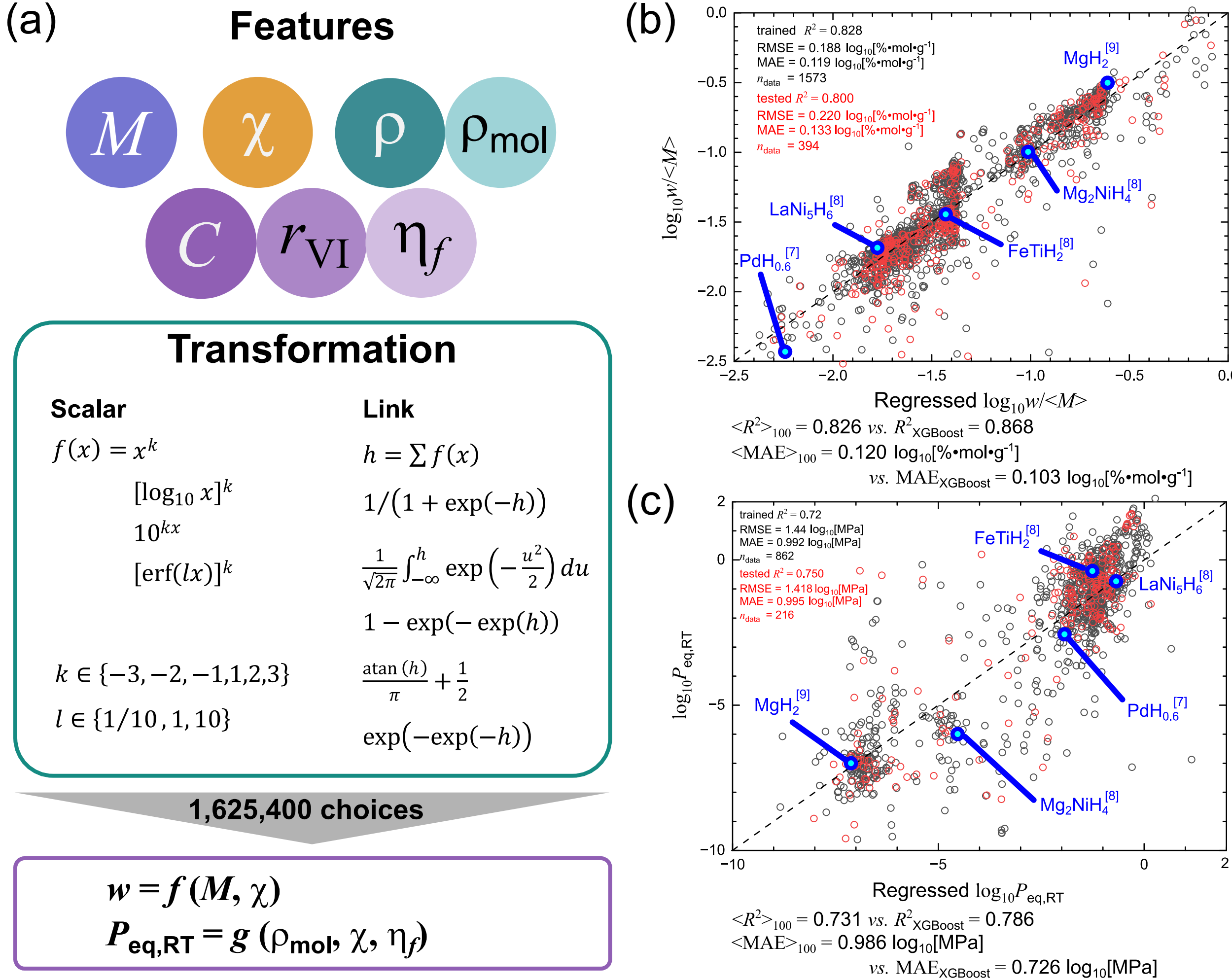

**Figure 3. Physically interpretable regression models for hydrogen storage properties.** (a) Seven elemental descriptors were combined with scalar transformations and (non)linear link functions to generate 1,625,400 candidate models. Final models for (b) gravimetric hydrogen density ($w$) [based on $\langle M \rangle$ and $\langle \chi - \chi_{\mathrm{H}} \rangle$; defined in Eq. (1)] and (c) equilibrium pressure ($P_{\mathrm{eq,RT}}$) at room temperature [$\langle \rho_{\mathrm{mol}} \rangle$, $\langle \chi - \chi_{\mathrm{H}} \rangle$, and $\langle \eta_f \rangle$; Eq. (2)], wherein training and test datasets follow an 80: 20 split and are denoted by black and red open circles, respectively. For both (b) and (c), 5 representative cases (*i.e.*, $MgH_2$, $Mg_2NiH_4$, $FeTiH_2$, $PdH_{0.6}$, and $LaNi_5H_6$),[7-9] the coefficient of determination ($R^2$) and mean absolute error (MAE) for the test datasets are provided, comparing with the ML benchmarks.[17]

As represented in **Figure 3a**, to enable physically interpretable modeling with low computational cost, we selected seven primary elemental features: the average atomic mass $\langle M \rangle$, the average electronegativity respect to hydrogen's $\langle \chi - \chi_{\mathrm{H}} \rangle$, the average density $\langle \rho \rangle$,[18] the average molar density $\langle \rho_{\mathrm{mol}} \rangle$, the average valence $\langle C \rangle$, the average Shannon ionic radius assuming sixfold coordination $\langle r_{\mathrm{VI}} \rangle$,[19] and the average ionic filling factor $\langle \eta_f \rangle$, of metal atoms (excluding hydrogen). More details can be found in the **Methods Section, Supporting Information**. The interrelationships among the 7 features are illustrated by a Pearson correlation heatmap (**Figure S2**). Furthermore, to capture the distributional characteristics of each feature, we included both the standard deviation $\sigma(x)$ and the skewness $r(x)$ for each descriptor, where *x* represents any of the following properties: $x = M, \chi - \chi_{\mathrm{H}}, \rho, \rho_{\mathrm{mol}}, C, r_{\mathrm{VI}}$, and $\eta_f$.

To comprehensively explore possible regression models from the pool of 21 candidate descriptors, 1,625,400 regression models in total were constructed and assessed by using multivariate linear regression modeling and multivariate beta regression modeling, which also enables nonlinear fitting.[20-24] More details can be found in the **Methods Section, Supporting Information**. The final models for $w$ (in the unit of %) and $P_{\mathrm{eq,RT}}$ (MPa) were selected based on those exhibiting the highest coefficient-of-determination $R^2$ values on test datasets, which were randomly sampled to comprise approximately 20% of the total data: 394 data points for $w$ and 216 for $P_{\mathrm{eq,RT}}$. Among the 1,625,400 candidate models evaluated, the optimal regression model for predicting $w$ was identified as a two-descriptor expression involving $\langle M \rangle$ and $\langle \chi - \chi_{\mathrm{H}} \rangle$ (**Figure 3b**). The final expression takes the following form:

$$\log_{10} \frac{w}{\langle M \rangle} = 3.78(1 - \exp[-\exp[2.60 - 1.62 \log_{10}\langle M \rangle - 0.234\langle \chi - \chi_{\mathrm{H}} \rangle]]) - 3.74. \qquad (1)$$

The regression model achieved $R^2 = 0.828$, root mean squared error RMSE $= 0.188 \log_{10} [\% \cdot \mathrm{mol} \cdot \mathrm{g}^{-1}]$, and mean absolute error MAE $= 0.119 \log_{10} [\% \cdot \mathrm{mol} \cdot \mathrm{g}^{-1}]$ for the training dataset (wherein the size of the dataset was given as $n_{\mathrm{data}} = 1573$) and $R^2 = 0.800$, RMSE $= 0.220 \log_{10} [\% \cdot \mathrm{mol} \cdot \mathrm{g}^{-1}]$, and MAE $= 0.133 \log_{10} [\% \cdot \mathrm{mol} \cdot \mathrm{g}^{-1}]$ for the test dataset ($n_{\mathrm{data}} = 394$). To further evaluate the model's robustness, we performed 100 random resamplings independently, each selecting $n_{\text{data}} = 394$ for testing. The resulting performance metrics were averaged to yield $\langle R^2 \rangle_{100} = 0.826 \pm .00101$, $\langle \text{RMSE} \rangle_{100} = 0.191 \pm 0.000374 \log_{10} [\% \cdot \text{mol} \cdot \text{g}^{-1}]$, and $\langle \text{MAE} \rangle_{100} = 0.120 \pm 0.0000483 \log_{10} [\% \cdot \text{mol} \cdot \text{g}^{-1}]$. For reference, our own XGBoost analysis on the same dataset, refined using 10-fold cross-validation (see **XGBoost Regression Section, Supporting Information**), yielded $R^2 = 0.868$, $\text{RMSE} = 0.164 \log_{10} [\% \cdot \text{mol} \cdot \text{g}^{-1}]$, and $\text{MAE} = 0.103 \log_{10} [\% \cdot \text{mol} \cdot \text{g}^{-1}]$ for the test data points, confirming that proposed descriptor-based model achieves accuracy on par with state-of-the-art ML approaches. Importantly, the model generalizes well across chemical space: 5 representative compounds (*i.e.*, $MgH_2$, $Mg_2NiH_4$, $FeTiH_2$, $PdH_{0.6}$, and $LaNi_5H_6$)[7-9] lie close to the "experimental = regressed" parity line, demonstrating both accuracy and practical utility in capturing real-world behavior.

We also provide detailed statistics of the regression model in **Table S1**, including standard errors, 95% confidence intervals, standardized coefficients, $t$-values, and variance inflation factors (VIF) for each term. All $t$-tests yield $p$-values $< 10^{-15}$, confirming that no term is redundant. The VIF value is suppressed (1.47), indicating the absence of multicollinearity. In addition, **Figure S3** presents histograms of the residuals for both training and test datasets, showing zero-centered distributions. This result demonstrates that the model errors are random rather than systematic, with no apparent bias or pattern.

The optimal regression model for predicting $P_{\text{eq,RT}}$ was derived using three physically meaningful descriptors: $\langle \rho_{\text{mol}} \rangle$, $\langle \chi - \chi_{\text{H}} \rangle$, and $\langle \eta_f \rangle$ (**Figure 3c**). The resulting model is expressed as:

$$\log_{10} P_{\text{eq,RT}} = 12.2 \left(1 - \exp\left[-\exp\left[-1.37 + 21.0 \langle \rho_{\text{mol}} \rangle - 0.163 \cdot 10^{-\langle \chi - \chi_{\text{H}} \rangle} - 0.878 \left[\text{erf}\left(10 \langle \eta_f \rangle\right)\right]^2\right]\right]\right) - 9.52. \quad (2)$$

The regression model achieved $R^2 = 0.728$, $\text{RMSE} = 1.44 \log_{10} [\text{MPa}]$, and $\text{MAE} = 0.992 \log_{10} [\text{MPa}]$ for the training dataset ($n_{\text{data}} = 862$) and $R^2 = 0.750$, $\text{RMSE} = 1.418 \log_{10} [\text{MPa}]$, and $\text{MAE} = 0.995 \log_{10} [\text{MPa}]$ for the test dataset ($n_{\text{data}} = 216$). To further evaluate the model's robustness, we performed 100 random resamplings independently, each selecting $n_{\text{data}} = 216$ for testing. The resulting performance metrics were averaged to yield $\langle R^2 \rangle_{100} = 0.731 \pm .00200$, $\langle \text{RMSE} \rangle_{100} = 1.42 \pm 0.0136 \log_{10} [\text{MPa}]$, and $\langle \text{MAE} \rangle_{100} =$ $0.986 \pm 0.00290\ \log_{10}$ [MPa]. For reference, our own XGBoost analysis on the same dataset, refined using 10 -fold cross-validation (see **XGBoost Regression Section, Supporting Information**), yielded $R^2 = 0.786$, $\text{RMSE} = 1.281\ \log_{10}$ [MPa], and $\text{MAE} = 0.726\ \log_{10}$ [MPa] for the test data points, further confirming that the descriptor-based model performs on par with advanced ML methods again. Notably, the same 5 reference compounds remain well-aligned with the "experimental = regressed" trend, reinforcing the robustness of the models across both thermodynamic and capacity domains.

Detailed regression statistics are provided in **Table S2**. All terms are significant ($p < 10^{-11}$), and VIF values below 3 indicate no multicollinearity. Residual histograms in **Figure S4** show zero-centered distributions, confirming that model errors are random rather than systematic. The use of simple, physically grounded descriptors thus enables predictive, interpretable modeling of hydrogen storage performance with broad chemical applicability. A complete listing of the feature vales $\chi - \chi_{\mathrm{H}}$, $M$, $\rho_{\mathrm{mol}}$, and $\eta_f$ is available in **Table S3**.

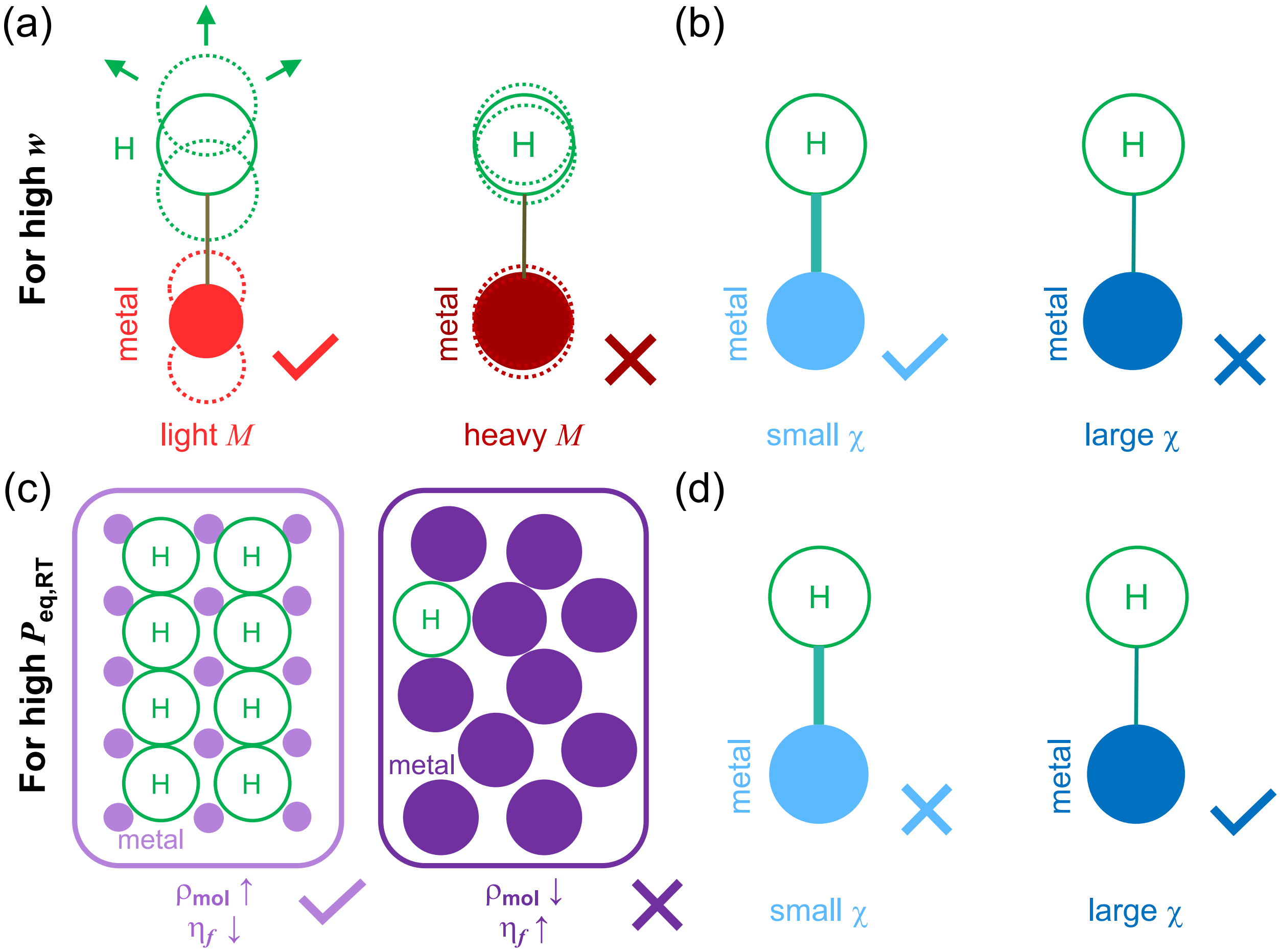

**Figure 4. Schematic interpretation of key descriptors influencing hydrogen storage performance.** High gravimetric density ($w$) is favored by: (a) light mass ($M$) and (b) large electronegativity ($\chi$) (strong bond polarity) of host metal atoms. High equilibrium pressure ($P_{\mathrm{eq,RT}}$) at room temperature is promoted by: (c) high molar density ($\rho_{\mathrm{mol}}$), low ionic filling factor ($\eta_f$), and (d) small $\chi$ (weak bond polarity).

To elucidate the physical basis of the regression models, we provide schematic illustrations of how each key descriptor influences hydrogen storage behaviors (**Figure 4**). For achieving high $w$, two primary factors are beneficial: a low $M$ and a large $|\langle \chi - \chi_{\mathrm{H}} \rangle|$ (given by small $\chi$; $\chi < \chi_{\mathrm{H}}$). As shown in **Figures 4a-b**, lighter host metals not only reduce the overall system mass, directly contributing to higher weight-specific capacity, but also may facilitate more dynamic lattice vibrations, potentially enhancing hydrogen diffusion into the bulk. A larger $|\langle \chi - \chi_{\mathrm{H}} \rangle|$ promotes stronger metal-hydrogen bond polarity, which is advantageous for maximizing hydrogen uptake under non-equilibrium or storage-focused conditions. Together, these descriptors encode essential atomic-scale design principles, providing both predictive power and chemical insights into the governing factors that enhance $w$ in metal hydrides.

For $P_{\mathrm{eq,RT}}$, the relevant descriptors reflect a different physical regime, rooted in thermodynamic stability. As illustrated in **Figures 4c-d**, high $\rho_{\mathrm{mol}}$ increases the number of reactive sites per volume, while a low $\eta_f$ implies reduced steric hindrance, allowing hydrogen to occupy interstitial positions more readily. Furthermore, smaller $|\langle \chi - \chi_{\mathrm{H}} \rangle|$, implying weaker bond polarity, lead to higher hydrogen chemical potential in the solid phase. This destabilizes the metal-hydrogen bond and shifts the equilibrium toward desorption, resulting in higher $P_{\mathrm{eq,RT}}$. Collectively, these effects enhance lattice accessibility and reduce hydrogen binding strength under equilibrium conditions, providing a chemically intuitive explanation for the model's predictive trends.

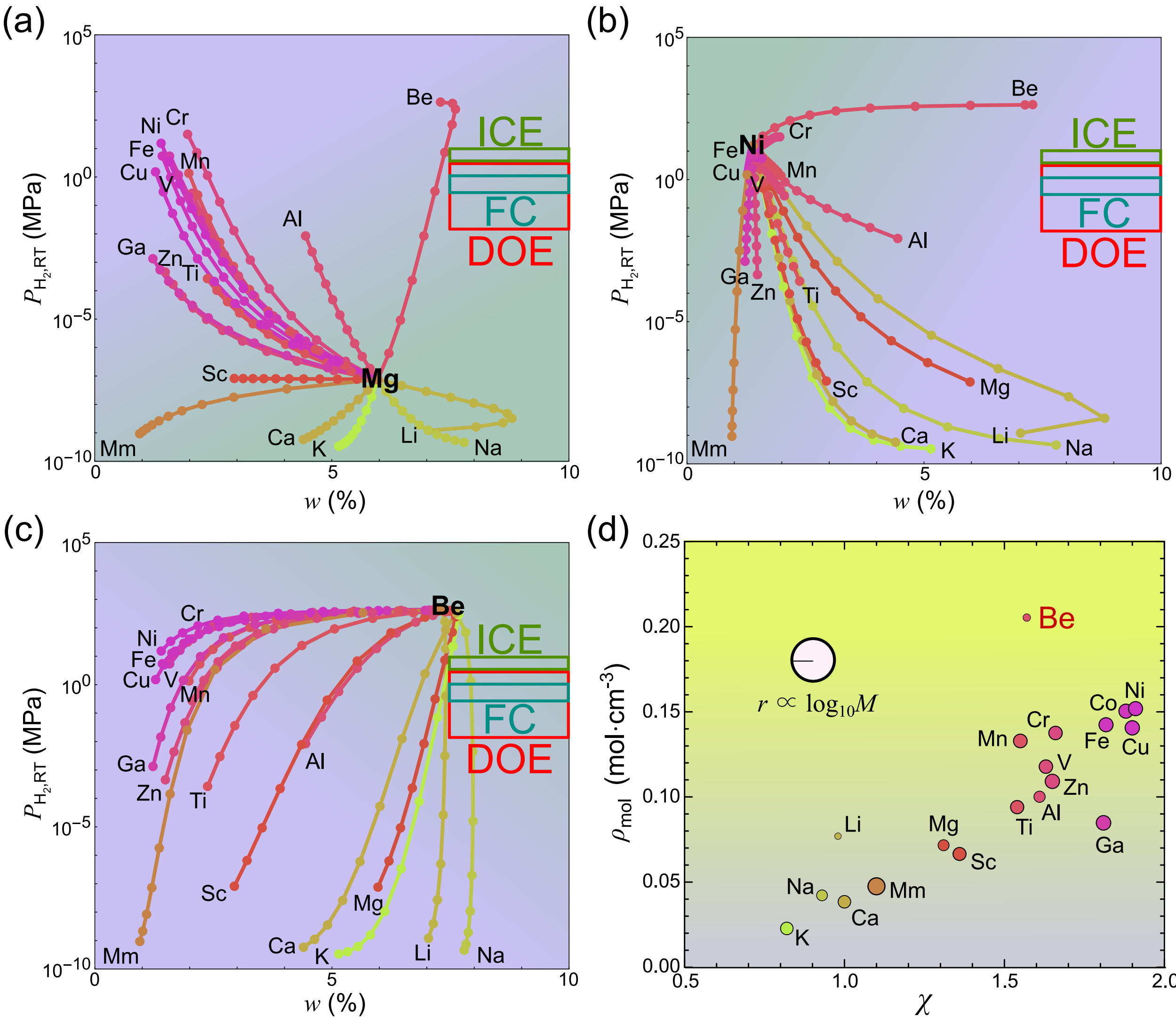

**Figure 5. Descriptor-based design maps and compositional pathways for hydrogen storage materials.** Predicted gravimetric density ($w$) and equilibrium pressure ($P_{\mathrm{eq,RT}}$) at room temperature for compositions interpolated between (a) Mg [(b) Ni] and other metals, tracing transitions from saline(interstitial)-type hydrides, respectively. (c) Compositional map anchored on Be, illustrating a unique trajectory that originates near the ultimate target regions of US-DOE. Three US-DOE benchmarks are indicated; red, dark green, and blue boxes represent the ultimate target converted from operating ambient temperature via van't Hoff relationships, that for internal combustion engine (ICE) applications, and that for fuel cell (FC) applications, respectively. Each dot in (a)-(c) represents a 10-atomic-percentage substitution step (*e.g.*, $Be \rightarrow Be_{0.1}Na_{0.9} \rightarrow \ldots \rightarrow Na$); color gradients follow electronegativity values $\chi$ as mapped in (d). (d) Distribution of elemental descriptors across metals: electronegativity ($\chi$, horizontal), molar density ($\rho_{\mathrm{mol}}$, vertical), and atomic mass ($M$, log-scaled as circle size).

To further explore how the regression models guide rational compositional design, we constructed descriptor-based design maps by selecting three representative elemental anchors (*i.e.*, Mg, Ni, and Be) and simulating compositional substitution trajectories with other metallic elements; additional compositional pathways originating from Li, Na, Al, K, Ca, Ce, Sc, Ti, V, Cr, Mn, Fe, Co, Cu, Zn, Ga, and Mm (mischmetal) are presented in **Figure S5**. Also, a user-interactive Excel file is provided in the **Supplementary Information** for predicting $w$ and $P_{\mathrm{eq,RT}}$ based on any input composition. As shown in **Figure 5a**, Mg, a prototypical saline-type hydride former, lies near the high-$w$, low-$P_{\mathrm{eq,RT}}$ corner of the map. Substituting Mg with other metals generally leads to increased $P_{\mathrm{eq,RT}}$ but reduced $w$, reflecting a shift away from the saline regime. Conversely, Ni-based pathways in **Figure 5b**, representing interstitial-type hydrides, exhibit the opposite trend: most substitutions lower $P_{\mathrm{eq,RT}}$ while increasing $w$. These two maps together highlight the $w$-$P_{\mathrm{eq,RT}}$ trade-off that characterizes conventional hydride systems.

In contrast, Be forms a unique trajectory on the design map, as represented in **Figure 5c**, with compositions such as $\mathrm{Be}_b\mathrm{Na}_{1-b}$. Three composition ranges, $b = 0.622\sim0.717$, $0.720\sim0.743$, and $0.673\sim0.698$ yield $w = 7.92\sim7.97\ \%$, $7.89\sim7.91\ \%$, and $7.93\sim7.95\ \%$, together with $P_{\mathrm{eq,RT}} = 0.016\sim3.1$ MPa, $3.5\sim10$ MPa, and $0.3\sim1.2$ MPa, respectively. These ranges correspond to the ultimate US-DOE targets converted from operating ambient temperature *via* the van't Hoff relationship (see **Ultimate US-DOE Targets of $P_{\mathrm{eq,RT}}$ *via* Van't Hoff Conversion Section, Supporting Information**), the targets for internal combustion engine (ICE) applications, and the targets for fuel cell (FC) applications, respectively.[25] No experimental reports on Be–Na hydrides are available to date. Be itself combines several rare and favorable features: low atomic mass ($M = 9.01\ \mathrm{g \cdot mol^{-1}}$), moderately high electronegativity ($\chi - \chi_{\mathrm{H}} = -0.63$), and the highest molar density ($\rho_{\mathrm{mol}} = 0.205\ \mathrm{mol} \cdot \mathrm{cm}^{-3}$) among the studied elements. These attributes together enable the balancing of $w$ and $P_{\mathrm{eq,RT}}$ that few other systems can offer.

To better understand the physical origin of the $w$-$P_{\mathrm{eq,RT}}$ trade-off, we analyze the elemental descriptor space in **Figure 5d**, where each metal is plotted by its $\chi$ and $\rho_{\mathrm{mol}}$, with $M$ encoded in circle size. A clear positive correlation is observed between $\chi$ and $\rho_{\mathrm{mol}}$, which may reflect that elements with stronger electron-attracting character favor shorter bond lengths and more compact lattice structures. This trend provides insight into the trade-off observed in **Figures 5a-b**; metals with low $\chi$ (*e.g.*, Mg, Li, Na) favor high $w$ due to comparably low $M$ and strong bond polarity to hydrogen (large $|\langle \chi - \chi_{\mathrm{H}} \rangle|$), but result in loose packing (low $\rho_{\mathrm{mol}}$) and low $P_{\mathrm{eq,RT}}$. In contrast, transition metals (e.g., Ni, Fe, Cr), with high $\chi$, exhibit tight packing (high $\rho_{\mathrm{mol}}$) and high $P_{\mathrm{eq,RT}}$ but reduced $w$ due to heavy $M$ and weaker polarity-driven uptake (small $|\langle \chi - \chi_{\mathrm{H}} \rangle|$).

Beryllium (Be), however, emerges as a distinct outlier in this landscape. Despite its relatively moderate electronegativity, Be possesses an unusually high $\rho_{\mathrm{mol}}$ and low $M$, positioning it in a sparsely populated upper-left region of the design space in **Figure 5d**. This anomaly may be attributed to its unique electronic structure: the low principal quantum number (2) of its valence electrons facilitates tight orbital overlap and strong core-valence attraction, enabling a compact atomic arrangement. This high packing efficiency enhances $P_{\mathrm{eq,RT}}$, while the light $M$ supports high $w$. As such, Be sits at a rare convergence point of design principles, suggesting that its alloys, particularly when judiciously combined with more electropositive elements, may serve as promising leads for next-generation hydrogen storage materials.[26-29] However, the use of Be raises important safety concerns due to its known toxicity, particularly in powder or nanoparticle form,[30] as well as its considerable cost. While the target metrics for hydrogen storage performance are favorable, any application of Be-based materials would require stringent handling protocols and careful risk-benefit evaluation.

Interestingly, the promising behavior of Be-based hydrides, particularly $\mathrm{Be_2Ti}$, is supported by experimental and computational findings that align closely with our model predictions. Mealand and Libowitz assumed that $0.1\ \mathrm{MPa} < P_{\mathrm{eq,RT}} < 15\ \mathrm{MPa}$ for $\mathrm{Be_2TiH_3}$ with $w = 4.4\ \%$.[26] Our regression models predict $w = 5.6\ \%$ and $P_{\mathrm{eq,RT}} = 171\ \mathrm{MPa}$, reasonably capturing both the high $w$ and a tendency toward elevated $P_{\mathrm{eq,RT}}$. More recently, Kim, Iwakiri, and Nakamichi conducted PCI measurements on $\mathrm{Be_2Ti}$ and assumed $P_{\mathrm{eq,RT}} \gg 13$ MPa, exceeding the upper limit of their apparatus.[27] Although the experimentally observed $w$ under 13 MPa was limited to 0.57 % (not reaching equilibrium), this was attributed to surface BeO formation, which impedes hydrogen uptake. Importantly, their first-principles calculations, which estimated accessible stable hydrogen sites in the lattice, yielded $w = 5.4\%$, closely corroborating both our prediction and earlier empirical data.[26] Taken together, both studies appear to have failed to achieve clear equilibrium conditions, most plausibly due to the intrinsically high $P_{\mathrm{eq,RT}}$ of the investigated Be-based hydride, which exceeded experimental constraints. In addition to $\mathrm{Be_2Ti}$, Mealand and Libowitz also studied $\mathrm{Be_2Zr}$, reporting the formation of a $\mathrm{Be_2ZrH_{2.3}}$ phase with $w = 2.1$ % at $P_{\mathrm{eq,RT}} = 13$ MPa.[26] Our model forecasts $w = 3.5\%$ and $P_{\mathrm{eq,RT}} = 40$ MPa, again indicating good semi-quantitative agreement. These converging observations reinforce the notion that Be-based systems reside in a favorable region of the design space, offering a rare combination of high $w$ and moderate-to-high $P_{\mathrm{eq,RT}}$, despite known practical challenges such as Be toxicity and oxide passivation. These real-world cases provide compelling validation for our model and emphasize the potential of Be-centered hydride chemistries.

**Summary**

In this work, we have developed and validated physically interpretable regression models for predicting two key performance metrics of metal hydrides: $w$ and $P_{\mathrm{eq,RT}}$. Leveraging a rigorously curated dataset (*DigHyd*) and a minimal set of chemically meaningful descriptors, we constructed explicit analytic models that match the predictive accuracy of state-of-the-art ML methods, while preserving full physical transparency. Design maps generated from these models revealed a fundamental trade-off between $w$ and $P_{\mathrm{eq,RT}}$ performances, rooted in opposing trends in elemental properties, particularly $\chi$. Saline-type hydrides, composed of light and more electropositive elements, tend to exhibit high $w$ due to strong metal-hydrogen bond polarity but suffer from low $P_{\mathrm{eq,RT}}$. In contrast, interstitial-type hydrides based on heavier, more electronegative transition metals show the opposite behavior. Amid this trade-off landscape, Be-based systems, such as Be-Na alloys, emerge as rare candidates capable of balancing both metrics, owing to its unique combination of low $M$ and high $\rho_{\mathrm{mol}}$ (possibly owing to the compact electronic structure). These findings offer chemically intuitive insights into the design principles governing hydrogen storage materials.

Beyond binary hydrides, the regression framework established here is readily extensible to more complex compositional spaces, including ternary and high-entropy alloy systems, as well as porous materials such as MOFs and COFs. The incorporation of additional physically motivated descriptors, or integration with first-principles methods, may further enhance the scope and fidelity of the models. Importantly, identifying chemically benign analogs to high-performing but toxic systems like Be-based compounds remains an urgent priority. More broadly, this descriptor-driven modeling strategy offers a scalable and interpretable platform for data-guided materials discovery, with potential applicability across diverse energy-relevant domains where structure-property relationships remain poorly understood. We also note that the present models are limited to equilibrium and gravimetric metrics; dynamic stability and cycling durability remain unaddressed and will be explored in future extensions of the framework.

## Associated content

Supporting Information: Methods, XGBoost regressions, ultimate US-DOE targets of $P_{\mathrm{eq,RT}}$ *via* Van't Hoff conversion, details of regression models, descriptor tables, and descriptor-based design map

Supporting Excel file: simulation wherein $w$ and $P_{\mathrm{eq,RT}}$ can be predicted for any user-defined composition

## Author information

## Data Availability

The supporting user-interactive Excel file (wherein $w$ and $P_{\mathrm{eq,RT}}$ can be predicted for any user-defined composition) is openly available in GitHub at https://github.com/gtex-project/dighyd-MLmodel. The data for ML modeling are stored in our *DigHyd* database (www.dighyd.org).

## Acknowledgments

This work was supported by The Green Technologies of Excellence (GteX) Program, Japan (Grant No. JPMJGX23H1).

**TOC**

**Interpretable models**

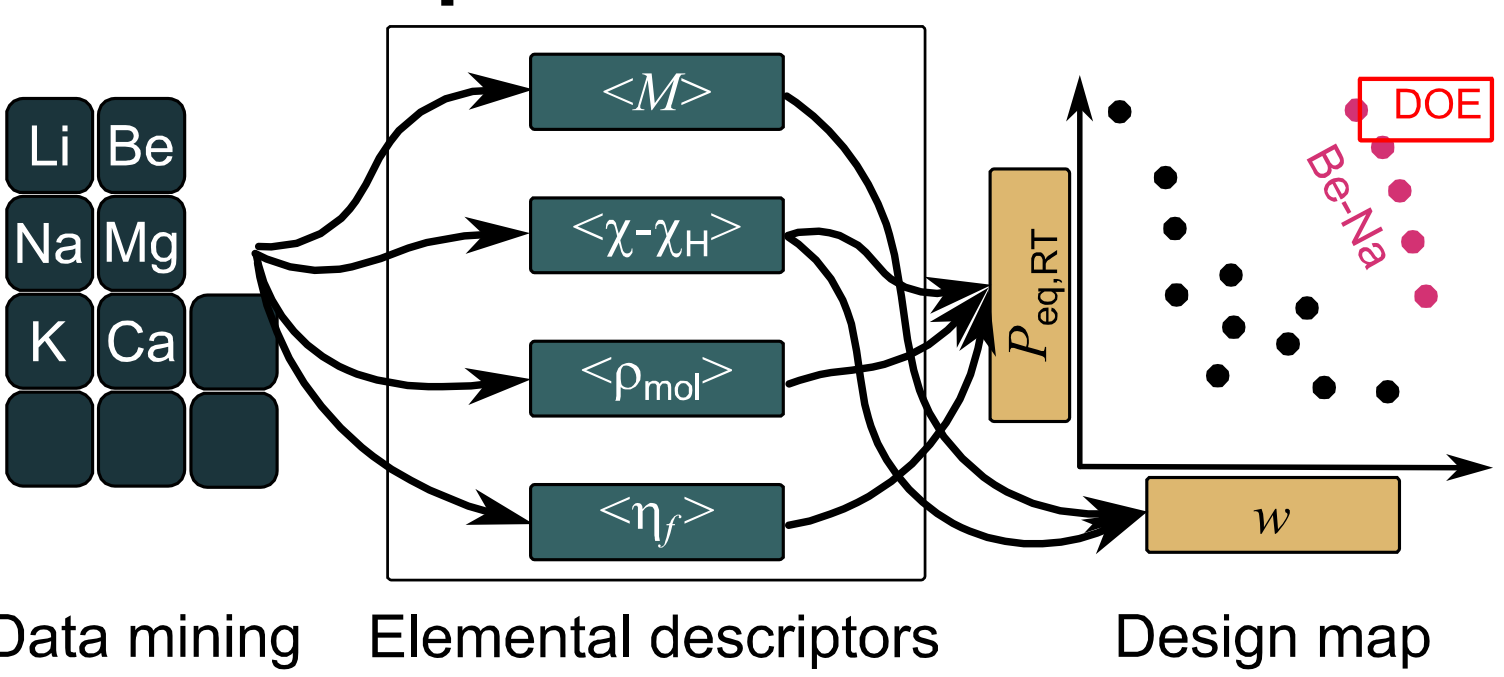